\documentclass[pra,aps,preprint,showpacs]{revtex4-2}
\usepackage[centertags]{amsmath}
\usepackage[utf8]{inputenc}
\usepackage{amsfonts}
\usepackage{amssymb}
\usepackage{amsthm}
\usepackage{amsmath}
\usepackage{newlfont}
\usepackage{epsfig}
\usepackage{amscd}
\usepackage{graphicx}
\usepackage{epsfig}
\usepackage{footnote}
\usepackage{lipsum}
\usepackage{color}
\usepackage{xcolor}
\usepackage{graphicx}
\usepackage{multirow}
\usepackage{enumerate}
\usepackage{caption}
\usepackage{subcaption}

\usepackage{amscd}

\newcommand{\beq}{\begin{equation}}
\newcommand{\eeq}{\end{equation}}

\newcommand{\ea}{\end{array}}
\newcommand{\bea}{\begin{eqnarray}}
\newcommand{\eea}{\end{eqnarray}}
\begin{document}

\begin{center}
{\large \sc \bf {Coherence restoring  in communication line via controlled interaction with environment}
}

\vskip 15pt

{\large
  E.B.~Fel'dman$^{1}$,  I.D.~Lazarev$^{1,2}$, A.N.~Pechen$^{3,4}$ and  A.I.~Zenchuk$^{1}$
}

\vskip 8pt

{\it $^1$
Federal Research Center of Problems of Chemical Physics and Medicinal Chemistry RAS,
Chernogolovka, Moscow reg., 142432, Russia}.

{\it $^2$
Faculty of Fundamental Physical-Chemical Engineering, Lomonosov Moscow State University, GSP-1, Moscow, 119991, Russia}.

{\it $^3$ Department of Mathematical Methods for Quantum Technologies, Steklov Mathematical Institute of Russian Academy of Sciences, Gubkina str. 8, Moscow 119991, Russia}

{\it $^4$National University of Science and Technology MISIS, Leninski prosp. 4, Moscow 119991, Russia}

{\it $^*$Corresponding author. E-mail:  zenchuk@itp.ac.ru}
\vskip 8pt

\end{center}
\date{\today}

\begin{abstract}
We consider the state-restoring protocol based on the controlled interaction of a linear chain with environment through { incoherent control by} the
specially adjusted  step-wise time dependent Lindblad operators. We show that the best restoring  result (maximal scale factors in the restored state) corresponds to the  symmetrical Lindblad equation. (0,1)-excitation dynamics is considered numerically, and restoring protocol  for the 1-order coherence matrix is proposed for the case of the two-qubit sender (receiver). The state-restoring with  equal scale factors is also considered reflecting the uniform scaling of the restored information.
\end{abstract}


\maketitle

\section{Introduction}
\label{Section:introduction}

Contemporary quantum information technology ~\cite{Schleich2016,Acin2018} motivates the  intensive development of quantum transmission protocols, which are based on photons \cite{PBGWK,PBGWK2,DLMRKBPVZBW} or, alternatively, on  spin chains \cite{PSB,LH} when dealing with  solid state quantum architectures.  The teleportation of arbitrary state \cite{BBCJPW,BPMEWZ,BBMHP}, arbitrary state transfer \cite{Bose,CDEL,KS,GKMT,GMT,ZASO,ZASO2,ZenchukJPA2012}, remote state creation~\cite{PBGWK,PBGWK2,DLMRKBPVZBW,PSB,LH,Z_2014,BZ_2015} represent different, but intertwined directions in development of this topic.
The intensive development of quantum technologies~\cite{Koch2022} stimulates further elaboration of optimization technique for quantum information transfer along communication lines.

\subsection{{ Brief review on quantum communication protocols}}
{ Among numerous results  devoted to quantum communication protocols we mention only several of them.  Quantum state transfer through one dimensional rings of qubits
with fixed  nearest neighbor   interactions constant throughout the ring is considered in Ref.\cite{Osborne2004}. 
Two parties are positioned at the opposite sites of the ring and both can operate with certain number of their nearest  qubits to increase the fidelity of state transfer. 
Fast high-fidelity information transfer  using 
the one-side control via  on-off switching actuator is considered in  Ref.\cite{SchirmerPRA2009}. The effective switching sequence can increase both speed and fidelity of state transfer, the particular form of the Hamiltonian has no principal role. 
The two side control of high-fidelity state transfer  along  the disordered spin chain governed by the nearest-neighbor Hamiltonian including only $x$-directed spin-spin interaction   is proposed in  Ref.\cite{Ashhab2015}. 
The state-transfer optimization tool in this case includes two Larmor frequencies at the end spins of the chain. 
The control of state transfer by the pair of time-dependent end-bonds in the XX-Hamiltonian with nearest-neighbor interaction is considered in Ref.\cite{OsendaPLA2021}.
The two-order magnitude increase in the rate of  state transfer  governed by  Heisenberg-type Hamiltonian can be reached via the optimal control of the  external parabolic magnetic field reaching  the theoretical maximum of propagation rate which is $\sim N/\langle D\rangle$ ($\langle D\rangle$ is the average coupling constant in the Hamiltonian),  as shown  in Ref.\cite{MurphyPRA2010}.
Speed limits for evaluation of one- two- and three- qubit operations  and   state transfer depending on the type of spin-interaction and geometry is studied in Ref.\cite{Ashhab2012}.
The protocol for preparing the multiqubit spin chain in the needed state using the optimized
coherent pulses, engineered dissipation and feedback is proposed in Ref.\cite{MorigiPRL2015}.
The state transfer along the odd-sized donor chain with realistic conditions is studied in 
Ref.\cite{Mohiyaddin2016}, where the nearest-neighbor  coupling constants and noise level are optimized for the high-fidelity state transfer. 
The spin dynamics and information transfer among the spins of the chain governed by either Heisenberg or Ising-Z -Hamiltonian where each spin is   subjected to a chaotic kicks is studied in Ref.\cite{AubourgJPB2016}. 
State transfer along the chain governed by the Heisenberg XXZ-Hamiltonian with periodic driving potential leads to either perfect state transfer or to complete blockade  of the input signal, Ref.\cite{ShanSciRep2018}. This feature can be used in formation of quantum logic elements. 
The  local control based on  non-adiabatic cutting and
stitching of end-bonds in  a spin chain using special time-dependent driving function can be also implemented in quantum information technology, Ref.\cite{PyshkinNJP2018}. 
In Ref.\cite{FerronPS2022}, 
it is demonstrated that almost perfect  single excitation state transfer is possible  using  a spin chain governed by  the Heisenberg Hamiltonian with either nearest-neighbor interactions, or involving remote spin interactions with optimized   site-dependent coupling constants. 
Non-trivial Floquet phases of matter are obtained  in the periodically driven Floquet quantum chain with special 
engineering, Ref.\cite{Peng2021}.
The control of a nuclear spin coherently coupled  with
the electron spins of  rare-Earth ions is studied in Ref. \cite{Uysal2023} in application to quantum repeater.  In this case the control is established by the special  sequence of magnetic pulses applied to  the electron spin.
The control via weakly coupled nuclear spins and electron  spins in the nitrogen-vacancy center in diamond is applied to multiqubit spin registers  in Ref.\cite{Taminiau2014} allowing to  prepare the high-fidelity single- and two-qubit gates  with error correction protocol. Many other aspects concerning controlling protocols in application to spin systems are explored in book~\cite{KuprovBook}.}

\subsection{{ Controlled interaction with environment}}
In this work, we exploit controlled interaction with the environment for coherence restoring in communication qubit chain lines. While dissipation and interaction with
the environment are often considered as having deleterious effect on the abilities to manipulate quantum systems, in some cases they can be exploited as a useful resource for controlling quantum systems { {as was shown in the approach of~\cite{PR_2006}, where a general incoherent control method for manipulation by quantum systems exploiting controlled interaction of the system with generally time-dependent environment and induced by this interaction time-dependent master equations was developed for various physical forms of the environment. In this approach, a quantum system interacts with a time-dependent environment and this interaction   leads to a master equation with time-dependent coefficients and with engineered dissipation. The master equation in particular can have the Gorini--Kossakowski--Sudarshan--Lindblad (GKSL) form but with time-dependent coefficients; however, non-Markovian master equations can be considered as well. Important that the time-dependent master equations in~\cite{PR_2006} were exploited not in the abstract form but in various explicit forms corresponding to various physical environments. Two  general types of the environment were considered as explicit examples---environment former by incoherent photons or phonons (Sec.~II in~\cite{PR_2006}) and collision type environments (e.g., gas, spin particles, etc.) which interact with the controlled system via collisions (Sec.~III in~\cite{PR_2006}). The first type of environment corresponds to the weak coupling and rotating wave Markovian approximation and the second type of the environment corresponds to the low density Markovian approximation. For both cases, explicit form of time-dependent Lindblad operators was written and exploited for quantum control. In this way, generally time-dependent state of the environment is used as a control to adjust the time-dependent decoherence rates and control evolution of the system density matrix. This incoherent control approach was shown to be quite powerful to provide, when combined with coherent control by lasers or electromagnetic field, creation of arbitrary mixed density matrices of generic quantum systems~\cite{P_2011}.} 

Controlled interaction with the environment and engineered dissipation were used { for engineering arbitrary pure and mixed quantum states of generic quantum systems~\cite{P_2011}} which are necessary} in some
models of quantum computation with mixed states and non-unitary 
gates~\cite{Aharonov, Tarasov2002, Verstraete2009, Schulte-Herbruggen}, in nuclear magnetic
resonance~\cite{LapertPRA2013}, steady quantum states generation \cite{science2015}, { for} improving quantum state engineering and computation~\cite{Verstraete2009}, cooling of
translational motion without reliance on internal degrees of freedom~\cite{Calarco2011}, preparing entangled states~\cite{Diehl_nature_2008,Weimer_nature_2010}, inducing
multi-particle entanglement dynamics~\cite{Barreiro_nature_2010}, making robust quantum memories~\cite{Pastawski_2011}, creating entanglement dynamics in two-qubit
systems~\cite{Ai_Fan_Jin_Cheng_2014}, and in other applications. Engineered interaction with the environment was used for making pointer-like states by suitably
engineering the coupling to the environment via open-loop periodic control~\cite{KhodjastehPRA2011}, dissipatively preparing arbitrary pure quantum states on a
multipartite qubit register in a finite number of basic control blocks~\cite{BaggioIEEE2012}, deterministically generating entanglement in an ensemble of strongly
interacting Rydberg atoms~\cite{RaoPRA2014}, preparing a spin chain in a generic many-body state in the asymptotic limit of tailored nonunitary
dynamics~\cite{MorigiPRL2015}, optimization of up-conversion hues in phosphors~\cite{LaforgeJCP2018}, preparation of qubit superposition states as the steady state of
coherent driving and dissipation near the photonic crystal band edge~\cite{HarringtonPRA2019}, {  manipulation by two-spin systems~\cite{MP_2023}}, driven-dissipatively preparing and stabilizing a quantum manybody entangled
state with symmetry-protected topological order~\cite{Wang2023}, { generation of target states and quantum gates in open quantum systems~\cite{PP_2022,PP_2023}}. Many more examples can be found in review~\cite{Harrington2022}.

\subsection{{ State-restoring algorithms}}
We study the  restoring of the
state transferred from the sender $S$ to the receiver $R$ along a spin-1/2 chain~\cite{FZ_2017,BFZ_Arch2018,Z_2018,FPZ_2021,BFLP_2022,FPZ_2024,BDFKLPZ_2024}.
The nodes of a spin chain connecting sender and receiver form the transmission line $TL$.
Initially, the  sender is
 prepared in the (arbitrary) state $\rho^{(S)}(0)$, while all the spins of $TL$ and $R$ are in the ground states. Then the prepared initial state is subjected to the time-evolution governed by some Hamiltonian,  effected by the  interaction with environment, up to some time instant
$t_{\mathrm{reg}}$ which we call time instant for state registration.
The initial sender state is restored if,  at that time instant,  the elements of the receiver's density matrix
 $\rho^{(R)}(t_{\mathrm{reg}})$ are  proportional
to the appropriate elements of the sender's initial density matrix  $\rho^{(S)}(0)$:
\begin{eqnarray}\label{restoring}
\rho^{(R)}_{ij}(t_{\mathrm{reg}})=\lambda_{ij}\rho^{(S)}_{ij}(0).
\end{eqnarray}
{ Here the $\lambda$-parameters $\lambda_{ij}$ do not depend on the initial state $\rho^{(S)}(0)$
and, generically, do not equal each other. Moreover,
Eq.~(\ref{restoring}) is supplemented  by  the trace-normalization  condition for the mixed quantum state, which constrains the structure of
the restorable sender's initial state $\rho^{(S)}(0)$~\cite{FPZ_2021,BFLP_2022}. Otherwise the restoring condition (\ref{restoring}) must be discarded at least for one of the diagonal elements of the density matrix.
Then our goal is to find the maximal by absolute value parameters $\lambda_{ij}$  in
Eq.~(\ref{restoring}). To this end,  it is sufficient to maximize the minimal among all $\lambda$-parameters.
The time instant $t_{\mathrm{reg}}$ corresponding to this maximum is the optimal time instant for state registration.

The state restoring protocol in \cite{FZ_2017,BFZ_Arch2018,Z_2018,FPZ_2021,BFLP_2022} relies on the special unitary transformation applied to the so-called extended receiver (receiver with several spins of the transmission line), which serves to provide the sufficient number of free parameters for restoring the transferred state. The larger extended receiver, the larger $\lambda$-parameters (by absolute value) can be achieved.

However, generating the required unitary transformation is not a simple task. Moreover,  although any unitary transformation can be constructed using one- and two-qubit operations according to the Solovay-Kitaev theorem \cite{KChV,NCh}, this construction is not effective in general. Therefore, alternative methods of state restoring are of practical significance.

In  \cite{FPZ_2024},    we modify the state restoring protocol by replacing the unitary transformation of the extended receiver with the time-dependent inhomogeneous magnetic field in the Hamiltonian. We require that this magnetic field acts selectively on some nodes of the spin chain and thus performs the goal of state restoring allowing to avoid introduction of the special unitary transformation at the receiver side. Thus, the local intensities of this magnetic field play the role of the parameters of the unitary transformation restoring the transferred state. In this case, the restoring parameters are encoded into the evolution operator rather than collected in the local unitary transformation, unlike Ref.~\cite{BFLP_2022}.  The developed protocol was applied to XX- \cite{FPZ_2024} and XXZ-model \cite{BDFKLPZ_2024}.

Now we study another state-restoring model based on the controlled interaction with environment \cite{PR_2006,MP_2023}.
We use the step-wise time dependence of the Lindblad operators \cite{PP_2022,PP_2023} with special values of jumps of control functions. Thus, the state-restoring parameters are included into the time-dependent Hamiltonian and can not be collected into the separate unitary operator, similar to the state-restoring via inhomogeneous time-dependent magnetic field \cite{FPZ_2024,BDFKLPZ_2024}.  Similar to that method, the obstacle for analytical representation of the parameter-dependent evolution operator exists in this case as well. However, we consider the relatively short chains that allows to overcome that obstacle without appealing to approximating the evolution operator via the Trotterization method \cite{Trotter,Suzuki}.

We shall emphasize that the investigation of restoring of quantum coherences has the potential to play a significant role in the physics of living systems~\cite{Chin2012}.
For instance, the investigation of the dynamics of quantum coherence in the chemical compass~\cite{Aiache2024} reveals
that the control by  the external magnetic field can be utilized to enhance and control the lifetime of entanglement.
In addition, the quantum coherences can act as a thermodynamic resource that enhances the efficiency of photoisomerization in coupled molecular systems
and hence pretend on potentially significant role in vision~\cite{Burkhard2024}.
At the same time, the decoherence as a consequence of interaction with environment is usually presented as substantial challenge.
On the contrary, in our paper we demonstrate that the particular turning of interaction with environment permits the quantum coherence restoring.

\subsection{{ Optimization methods}} 
In the following, we give a brief overview of available optimization methods that can be used to achieve the purpose of maximizing the $\lambda$-parameters in Eq.(\ref{restoring}).

A popular method used for optimization of control parameters in spin dynamics is GRadient Ascent Pulse Engineering (GRAPE) approach which originally was developed for optimizing the NMR pulse sequences for control of quantum quantum dynamics~\cite{Khaneja2005}.  Later, this method was applied to optimization problems in various systems~\cite{GlaserJPB2011,Fouquieres2011,PechenTannor2011,Lucarelli2018,PP_2023}. In particular,  GRAPE approach was used in analysis of quantum dynamics governed by a Hamiltonian with coherent control in~\cite{VMP_21} and quantum dynamics involving spins interaction  with the environment and thus driven by both coherent control and  environmental control~\cite{PP_2023}. Open GRAPE allows the high-fidelity  realization of a CNOT ~\cite{GlaserJPB2011}. However, applying GRAPE approach to large spin systems is still to be developed because the large dimensionality creates a real obstacle for effectiveness of this method.

Among the various alternative approaches,
the Chopped Random Basis (CRAB) optimization method has gained popularity
due to its ability to reduce the dimensionality of the control problem by
expanding the control field on a truncated basis of random functions~\cite{Doria2011}.
This characteristic renders CRAB particularly well suited for
experimental implementations where the number of adjustable control
parameters is limited.
The method has been successfully applied in various quantum control
scenarios, including optimal control in closed and open quantum systems~\cite{Koch2022,Muller2022}.
However, despite its advantages in simplifying the optimization
landscape, the implementation of CRAB can become complex when high
precision and fast convergence are required, especially in
high-dimensional spin systems~\cite{Muller2022ScientificReports}.

In this paper, we present a solution to the quantum state-restoring problem using the least squares method with a regularization functional.
This approach enables us to achieve optimal results.

\vspace{0.5cm}

The paper is organized as follows. In Sec.~\ref{Section:T}, we consider the general state-restoring protocol based on the time-depending Lindbladian. Evolution under the time-dependent Lindbladian including XXZ-Hamiltonian is discussed in Sec.~\ref{Section:ev}. State restoring of (0,1)-excitation states with numerical simulation of 8 and 10 node chains and two-qubit sender (receiver) is given in Sec.~\ref{Section:01}. Concluding remarks are presented in Sec.~\ref{Section:conclusions}.

\section{General state restoring protocol}
\label{Section:T}
We consider the { Born-Markov approximation for evolution of a quantum system interacting with the environment is described by the master equation of the general GKSL form
\begin{eqnarray}\label{rho00}
 \rho_t(t)&=&-i (H\rho(t) - \rho(t) H) +\\\nonumber
&& \sum_{i=1}^N \Big(L_i(t) \rho(t) L_i^\dagger(t)
-\frac{1}{2} L_i^\dagger (t) L(t) \rho(t) -\frac{1}{2} \rho(t) L_i^\dagger(t) L(t)\Big).
\end{eqnarray}
{ A rigorous derivation of the general form of Markovian master equations was performed in Refs.\cite{Gorini1976,Lindblad,Carmichael}. For a simple physical justification one can look, for example in \cite{Preskill}. Time-dependent form of the master equation with explicit forms of the time-dependent Lindblad operators $L_i(t)$ corresponding to two physical models of the environment, namely environment formed by photons or phonons (corresponding to weak coupling and rotating wave markovian approximation), and environment formed by collisional type media (quantum gas, spin systems, etc., corresponding to low density approximation) was considered in~\cite{PR_2006}. 

We consider an additional requirement that} both the Hamiltonian $H$ and the GKSL operators $L_i$ conserve the excitation number in the system.  A particular realization of such Lindblad operators is proposed below in Sec. \ref{Section:ev}. The requirement on conserving the excitation number  is auxiliary, it is included to simplify description of the spin dynamics. In our case, this dynamics will be described by two  block of the  Lindblad operator responsible for 0- and 1-excitation  evolution. In general,  Eq.(\ref{rho00}) holds for any Hamiltonian and Lindblad operators. 
} We emphasize that the Lindblad operators are time-dependent which is necessary to establish the control over the restoring process. The state restoring protocol is similar to that proposed in \cite{FPZ_2024}, but there is significant difference because of  the non-Hamiltonian evolution.

Our one-dimensional communication line includes the sender $S$ {($N^{(S)}$ nodes)}, receiver $R$ {($N^{(R)}=N^{(S)}$ nodes)} and transmission line $TL$  {($N^{(TL)}$ nodes)} connecting them.
We are aiming at solving the initial value problem with the
initial state
\begin{eqnarray}\label{rho0}
\rho(0)=\rho^{(S)}(0)\otimes \rho^{(TL;R)}(0),
\end{eqnarray}
where $\rho^S(0)$ is an arbitrary initial sender's state to be transferred to the receiver $R$, while the initial state of the transmission line and receiver $\rho^{(TL;R)}(0)$ is the ground state,
\begin{eqnarray}
\rho^{(TL;R)}(0) = |0_{TL,R}\rangle
{\langle 0_{TL,R}|} = {\mbox{diag}}(1,0,\dots).
\end{eqnarray}
Then, formally, we can represent the evolution of the density matrix $\rho(t)$ in the following element-wise form:
\begin{eqnarray}\label{rhonm}
\rho_{nm}(t)=\sum_{{i,j=0}\atop{|i|=|n|, |j|=|m|} }^{N^{(S)}-1} U_{nm;ij}(t)\rho_{ij}^{(S)}(0) ,\;\;n,m=0,\dots,2^N-1,
\end{eqnarray}
where $|n|$ is the number of units in the binary representation of the index $n$,
 $U(t)$ is  the evolution operator defined by the Lindblad equation (\ref{rho00})
with the initial condition
\begin{eqnarray}\label{ic}
U_{nm;ij}(0) =
 \delta_{ni}\delta_{mj},\;\;i,j,n,m=0,\dots,2^{N}-1.
\end{eqnarray}
Note that evolution conserving the excitation number in the system induces the following constraints for the subscripts of the operator $U_{nm;ij}$:
\begin{eqnarray}
U_{nm;ij}: \;\;|i|=|n|,\;\;|j| = |m|,
\end{eqnarray}
which are reflected in the summation limits in (\ref{rhonm}).

The Hamiltonian preserving the excitation number must commute with the $z$-projection of the total spin momentum $Z=\sum Z_i$, $Z_i$ is the operator of $z$-projection of the $i$th spin momentum, $Z_i =\frac{1}{2}{\mbox{diag}}(1,-1)$,
\begin{eqnarray}\label{com0}\label{com}
[H,Z]=0.
\end{eqnarray}
In turn, commutation relation (\ref{com0}) imposes the block-diagonal structure on  the Hamiltonian written in the basis with ordered number of excitations,
\begin{eqnarray}\label{bH}
H={\mbox{diag}} (H^{(0)},H^{(1)},\dots ),
\end{eqnarray}
where the block $H^{(j)}$ governs the evolution of the $j$-excitation subspace and $H^{(0)}$ is a scalar.
The same diagonal block-structure is required for $L_i$:
\begin{eqnarray}\label{bH2}
L_i={\mbox{diag}} (L_i^{(0)},L_i^{(1)},\dots ), \;\; i=1,\dots,N.
\end{eqnarray}
We consider the restoring problem for the  nondiagonal elements of the receiver density matrix $\rho^{(R)}$,
\begin{eqnarray}\label{R}
\rho^{(R)}={\mbox{Tr}}_{S,TL}(\rho).
\end{eqnarray}
Calculating the partial trace in (\ref{rhonm}) we
 split subscripts $n$, $m$, $i$ and $j$ as follows
\begin{eqnarray}\label{SR}
n\to (n_{S,TL} n_R), \;\;m\to (n_{S,TL} m_R),\;\;i\to (i_S 0_{TL,R}),  \;\;j\to (j_S 0_{TL,R}),
\end{eqnarray}
thus selecting nodes of  the receiver and sender.
Then
\begin{eqnarray}
\rho^{(R)}_{n_Rm_R} =\sum_{n_{S,TL}} \sum_{i_Sj_S}U_{(n_{S,TL} n_R)(n_{S,TL}m_R);(i_S0_{TL,R})(j_S0_{TL,R})} \rho^{(S)}_{i_Sj_S}(0)  =\sum_{i_Sj_S} T_{n_Rm_R;i_Sj_S}  \rho^{(S)}_{i_Sj_S}(0)  ,
\end{eqnarray}
where
\begin{eqnarray}\label{constr00}
T_{n_Rm_R;i_Sj_S} =\sum_{n_{S,TL}}U_{(n_{S,TL} n_R)(n_{S,TL}m_R);(i_S0_{TL,R})(j_S0_{TL,R})} .
\end{eqnarray}
Imposing the restoring conditions,
\begin{eqnarray}\label{restor1}\label{constr0}
T_{n_Rm_R;i_Sj_S} =0, \;\;(i_S,j_S)\neq (n_R,m_R),\;\;  n_R\neq m_R ,
\end{eqnarray}
we obtain
\begin{eqnarray}\label{rest}
\rho^{(R)}_{n_Rm_R}(t_{\mathrm{reg}}) = \lambda_{n_Rm_R} \rho^{(S)}_{n_Rm_R}(0), \quad n_R\neq m_R,
\end{eqnarray}
with
\begin{eqnarray}\label{lambda1}
\lambda_{n_Rm_R} =T_{n_Rm_R;n_Rm_R}, \;\; n_R\neq m_R.
\end{eqnarray}
We shall emphasize that the scale parameters $\lambda_{n_Rm_R}$ are universal, i.e., they do not depend on the elements of  the initial sender's density matrix and are completely determined by the Lindbladian  and the selected time instant for state registration {$t_{\mathrm{reg}}$}.

\section{State restoring via  Lindbladian  with XXZ-Hamiltonian}
\label{Section:ev}
We consider the evolution under the  Lindbladian with XXZ-Hamiltonian { and such  interaction with the environment that results only in  dephasing in the spin system. This type of  interaction can be described by the Lindblad operators 
$L_i(t) = \sqrt{\gamma_i(t)} Z_{i}$.} Since $Z_{i}^2=\frac{E}{4}$ ($E$ is the identity operator), Eq.(\ref{rho00}) gets the form
 \begin{eqnarray}\label{rho}
&& \rho_t(t)=-i (H  \rho(t) - \rho(t) H)+ \sum_{i=1}^N \gamma_i(t)\Big(Z_{i} \rho(t) Z_{i}
-\frac{1}{4} \rho(t) \Big),
\\\nonumber
&&H=\sum_{j>i} D_{ij} (X_{i}X_{j}+Y_{i}Y_{j}  -2 Z_i Z_j),\;\;
X_i=\frac{1}{2}\left(\begin{array}{cc}
0&1\cr
1&0
\end{array}\right),\;\;  Y_i=\frac{1}{2}\left(\begin{array}{cc}
0&-i\cr
i&0
\end{array}\right) .
\end{eqnarray}
{Here $D_{ij}=\gamma^2/(2r_{ij}^3)$ are the coupling constants between the $i$th and $j$th spins (for $\hbar=1$), $\gamma$ is the gyromagnetic ratio,
$r_{ij}$  is the distance between the $i$th and $j$th spins}, and {$X_i,\;\;Y_i$ are  the operators of, respectively, the $x$- and $y$-projections of the $i$th spin,} {the external magnetic field is { perpendicular to} the spin chain.} Of course, commutation condition (\ref{com}) is satisfied. 
{ We note that the rotating-wave approximation \cite{Abragam}, which was  used in deriving the XXZ-Hamiltonian,  is represented  by the unitary transformation $e^{i \gamma B I_z}$ ($B$ is the intensity of the external magnetic filed)  commuting with our Lindblad operators and therefore does not  change them.}

The block-structure of $L_i$ in (\ref{bH2}) is related to the block structure of $Z_i$:
\begin{eqnarray}
Z_i ={\mbox{diag}}(Z_i^{(0)}, Z_i^{(1)},\dots),
\end{eqnarray}
where
\begin{eqnarray}
Z_i^{(0)} =\frac{1}{2}, Z_i^{(1)} = \frac{1}{2} {\mbox{diag}}( \underbrace{1,\dots,1}_{i-1}, -1, \underbrace{1,\dots,1}_{N-i}, \;\;\dots.
\end{eqnarray}

To satisfy constraints (\ref{constr0}), we introduce the set of $N^{(ER)}$ ($ER$ means  Extended Receiver) nonzero controls
$\gamma_k(t)$, $k=N-N^{(ER)}+1, \dots, N$,
assuming
\begin{eqnarray}
\gamma_k=0,\quad  k=1,\dots,N-N^{(ER)}.
\end{eqnarray}
The nonzero $\gamma_k(t)$ are some functions of $t$. To simplify further analysis, let them be step-functions~\cite{PP_2023}
\begin{eqnarray}\label{LF}
\gamma_k(t) = \sum_{j=1}^{K_\gamma} a_{kj} \theta_j(t),\quad  \theta_j(t) =\left\{\begin{array}{ll}
1, &t_{j-1} < t\le t_j\cr
0  & {\mbox{otherwise}}. \end{array}
\right., \;\;k=N-N^{(ER)}+1,\dots,N.
\end{eqnarray}
Here we split the entire time interval
$[0,t_{\mathrm{reg}}]$
in $K_\gamma+1$ intervals of different (in general) lengths assuming that over the first interval $0\le t\le t_0\equiv t_{\mathrm{reg}}- \sum_{k=1}^{K_\gamma} \Delta t_k$, $\Delta t_k = t_k - t_{k-1}$, the evolution is governed by the $XXZ$-Hamiltonian without Lindblad terms (i.e., all  $\gamma_i=0$).
As a simplest variant, let us fix $t_j$ and consider $a_{kj}$ as control parameters.
Therefore, the extended receiver with $N^{(ER)}$ nodes has $K_\gamma N^{(ER)}$ free parameters.

We can write (\ref{rho}) in the  element-wise  form
\begin{eqnarray}
\label{rhonm2}
&&
\partial_t\rho_{nm}(t) = \sum_{ij} u^{(l)}_{nm;ij}\rho_{ij}(t), \;\; t_{l-1} < t \le t_l,\;\; l=1,\dots,K_\gamma\\\nonumber
&&
u^{(l)}_{nm;ij}=\left\{
\begin{array}{ll}
\displaystyle -i (H_{ni} - H_{jm}) +&\cr \displaystyle\sum_{k=N-N^{(ER)}+1}^N a_{kl} ((Z_k)_i (Z_k)_j \delta_{ni}\delta_{mj} - \frac{\Gamma_l}{4} \delta_{ni}\delta_{mj},& t_{l-1}<t\le t_l\cr
1,& {\mbox{otherwise}}\end{array}\right.,
\end{eqnarray}
where $\displaystyle\Gamma_l = \sum_{k=N-N^{(ER)}+1}^N a_{kl}$. Thus,   $u^{(l)}$ is constant over each  time-interval $t_{l-1}<t<t_l$.
 Let us introduce the matrix $\hat u^{(l)} =\{u^{(l)}_{nm;ij}\}$,  where each pair of indexes $(nm)$ and $(ij)$ is treated as a single index. Similarly, each element $\rho_{nm}$ is considered as an element of a vector $\vec{\rho}$.
Then Eq.(\ref{rhonm2}) can be integrated to result in
 \begin{eqnarray}\label{rhoT}
 \vec{\rho}(t) = U^{(l)}(t) \vec{\rho}(t_{l-1}),\;\; U^{(l)}(t)=e^{\hat u^{(l)} t}, \;\; t_{l-1} < t \le t_l.
 \end{eqnarray}
Thus, the operator $U$ in (\ref{rhonm})  becomes a product of piece-wise $N^2 \times N^2$ constant operators:
\begin{eqnarray}\label{UtK}
U(t_{\mathrm{reg}})&=&  U^{(K_\gamma)}(\Delta t_{K_\gamma})\dots U^{(1)}(\Delta t_1) U^{(0)}\Big(t_{reg} - \sum_{j=1}^{K_\gamma} \Delta t_j\Big) ,\quad \\\nonumber
&& \Delta t_j = t_j -t_{j-1},\\\label{W0}
U^{(0)}_{nm;ij}(t) &=&\left( e^{-iHt} \right)_{ni}
\left( e^{iH t} \right)_{jm}.
\end{eqnarray}
Although formally implementing  the evolution with  piece-constant $\gamma_j$  (\ref{LF}) is very simple,
its simulation faces the problem of  Hamiltonian diagonalization because of the free parameters $a_{kj}$, which can not be fixed until the restoring system (\ref{constr0}) is solved {and therefore must be treated symbolically}.
However, for comparably short chains this obstacle can be overcome {via contemporary computation technique.}

\section{
 Restoring  1-order coherence matrix}
\label{Section:01}
\label{Section:ord1}
Now we consider the (0,1)-excitation { state-space,} { i.e., the state-space spanned by the eigenvectors with 0- and 1 excitations,} and construct the protocol for restoring the elements of the 1-order coherence matrix $\rho^{(1)}$ only, thus leaving all elements of the 0-order coherence matrix unrestored.
In this case Eq.(\ref{rho}) gets the following form:
 \begin{eqnarray}\label{rho1}
&& \rho^{(1)}_t(t)=-i   \rho^{(1)}(t) (H^{(0)}-H^{(1)})+ \sum_{i=1}^N \gamma_i(t)\Big(\rho^{(1)}(t)Z^{(0)}_{i}  Z ^{(1)}_{i}
-\frac{1}{4} \rho^{(1)}(t) \Big).
\end{eqnarray}
Here, the matrix $\rho^{(1)}$ is the row of $N$ elements, $H^{(0)}$ and $Z_i^{(0)}$ are scalars, $H^{(1)}$ is a matrix $N\times N$ and $Z_i^{(1)}$ is a diagonal matrix $N\times N$.
Eq.(\ref{rhonm2}) can be written in the matrix form:
\begin{eqnarray}
\label{rhonm22}
&&
\partial_t\rho^{(1)}(t) =\rho^{(1)}(t) u^{(l)}, \;\; t_{l-1} < t \le t_l,\;\;l=1,\dots,K_\gamma,\\\nonumber
&&
u^{(l)}=\left\{
\begin{array}{ll}
\displaystyle -i (H^{(0)}- H^{(1)}) +&\cr\displaystyle \sum_{k=N-N^{(ER)}+1}^N a_{kl} ((Z_k^{0)}) (Z^{(1)}_k)  - \frac{\Gamma_l}{4},& t_{l-1}<t\le t_l\cr
1& {\mbox{otherwise}}\end{array}\right..
\end{eqnarray}
Eq.(\ref{UtK}) gets the following form:
\begin{eqnarray}\label{UtK1}
U(t_{\mathrm{reg}})&=&  U^{(0)}\Big(t_{reg} - \sum_{j=1}^{K_\gamma} \Delta t_j\Big)U^{(1)}(\Delta t_1)  \dots U^{(K_\gamma)}(\Delta t_{K_\gamma}) ,\quad \\\nonumber
&&
\end{eqnarray}
 where $U^{(k)}=e^{u^{(k)} t}$, $k=1,\dots, K_\gamma$, $U^{(0)}=e^{-i(H^{(0)}-H^{(1)})t}$. We emphasize that the RHS of (\ref{UtK1}) contains the product of usual $N\times N$ matrices.
Thus, using subscripts in the form (\ref{SR}), we rewrite Eq.(\ref{constr00}) for the elements of the 1-order coherence matrix as
\begin{eqnarray}
T_{m_R;j_S}(\gamma) \equiv T_{0_Rm_R;0_Sj_S}(\gamma)  =U_{j_S0_{TL,R};0_{S,TL}m_R}(\gamma),
\end{eqnarray}
where $\gamma = \{a_{jk}\}$ is the list of all free parameters in the Lindbladian.
The restoring conditions (\ref{constr0}) yield
\begin{eqnarray}\label{restoring1}
 T_{m_R,j_S}(\gamma)=0, \;\;j_S\neq m_R .
\end{eqnarray}
Finally, for the $\lambda$-parameters  (\ref{lambda1}) we have
\begin{eqnarray}\label{lambda12}
\lambda_{m_R} =T_{m_R,m_R}.
\end{eqnarray}

Let $\tilde \gamma = \{\tilde a_{jk}\}$ be the list of parameters solving system (\ref{restoring1}).
The solution to this system is not unique, we denote the $m$th solution by
$\tilde \gamma^{(m)}=\{\tilde a^{(m)}_{jk}\}$.
As the characteristic of the quality of the state restoring we use the parameter $\lambda$ (transmission quality)
\begin{eqnarray}\label{lammax}
\lambda = \max_{m} \min_{m_R} \{ |\lambda_{m_R}(\tilde \gamma^{(m)} )|
\},
\end{eqnarray}
($\lambda=1$ in the perfect case). We denote by
$\gamma^{\mathrm{(opt)}}$
the list of parameters $\tilde \gamma^{(m)}$ found as the result of maximization in (\ref{lammax}).
{ We emphasize that the parameter $\lambda$ in Eq.(\ref{lammax}) indicates how much the  absolute values  of the matrix elements of $\rho^{(S)}(0)$ are decreased during the state transfer and restoring. This is the most important characteristics of restoring process because vanishing $\lambda$ means loosing the transferred information. However, the phases of the $\lambda$-parameters  are also important because they are added to the phases of the elements of $\rho^{(S)}(0)$. Therefore, both absolute values and phases of  the $\lambda$-parameters must be kept in the state-restoring protocol, see Refs.\cite{FZ_2017,BFZ_Arch2018,Z_2018,FPZ_2021,BFLP_2022}.}

\subsection{Numerical simulations, $N^{(S)}=2$}
In numerical simulations, we use the dimensionless time $\tau = D_{12} t$ and equal intervals
$\Delta\tau = \tau_j-\tau_{j-1}$. 
In this case, there are two equations in (\ref{restoring1}):
\begin{eqnarray}\label{U1002}
T_{01,10}(\gamma)=\delta_1 (\gamma)=0,\;\;\; T_{10,01}(\gamma)=\delta_2(\gamma)=0,
\end{eqnarray}
and two $\lambda$-parameters $\lambda_{01}$, $\lambda_{10}$, i.e., Eq. (\ref{lammax}) for the transmission quality $\lambda$ gets the form
\begin{eqnarray}\label{lammaxEx}
\lambda = \max_m \min \{ |\lambda_{01}(\tilde \gamma^{(m)} )|,  |\lambda_{10}(\tilde \gamma^{(m)} )|
\}.
\end{eqnarray}
We take $K_\gamma=3$ in (\ref{LF}), i.e. we split the entire time interval into three subintervals with
$\gamma_k=a_{kj}$ in the $j$th interval, $j=1,2,3$.
We set $N^{(ER)}=N$, i.e., $\tau_{\mathrm{reg}} - \sum_{j=1}^{K_\gamma} \Delta \tau_j=0$, so that the argument of $U^{(0)}$ in (\ref{UtK1}) is zero. {The last assumption is motivated by the preliminary simulations which show that the best restoring result corresponds to the earliest switching on the Lindblad terms. Thus, $\gamma=\{a_{kj},\; k=1,\dots,N,\; j=1,2,3\}$.}

The problem~(\ref{U1002}) is formally an ill-posed nonlinear system in the sense that
 its solution is not unique.
Solving this system via variational regularization allows one to find the approximate solutions
that are not only consistent with the original equations  (\ref{U1002}),
but also satisfy additional special conditions imposed through the regularization functional.
This ensures the selection of solutions possessing certain desirable properties.
Since we are interested in the solution for which  $\lambda$-parameters are maximized,
the regularization functional can be taken in the form%
 { \begin{eqnarray}\label{eq:U1002-regularization}
 R_\mathrm{\lambda} (\gamma)&=& \left(1 - |{\mbox{Re}}(\lambda_{01}( \gamma ))|\right)^2+\left(1 - |{\mbox{Im}}(\lambda_{01}( \gamma ))|\right)^2+ \\\nonumber
 &&
  \left(1 - |{\mbox{Re}}(\lambda_{10}( \gamma ))|\right)^2+\left(1 - |{\mbox{Im}}(\lambda_{10}( \gamma ))|\right)^2.
\end{eqnarray}}
%
%
Thus, we solve the system~(\ref{U1002}) 
with the regularization functional~(\ref{eq:U1002-regularization})
using the least-squares method
implemented in  the \texttt{least\_squared} function from the SciPy package~\cite{scipy}.
The sum of squared residuals { $\delta_1^2$ and $\delta_2^2$ and weighted regularization function} has the form
\begin{equation}\label{eq:target-function}
  S(\gamma) = \delta_1^2 (\gamma) + \delta_2^2 (\gamma)
  + \mu {   R_\mathrm{\lambda} (\gamma)}
\end{equation}
where $\mu$ is a parameter that controls the importance of regularization.
In all calculations presented below, we used $\mu = 10^{-6}$.
Finally, we use the approximate solution obtained at this stage as an initial approximation
to find the solution of the restoring system~(\ref{U1002}) without regularization functional (\ref{eq:U1002-regularization}), {i.e., setting $\mu=0$ in (\ref{eq:target-function}).}

\begin{figure}[ht]
  \includegraphics[width=\textwidth]{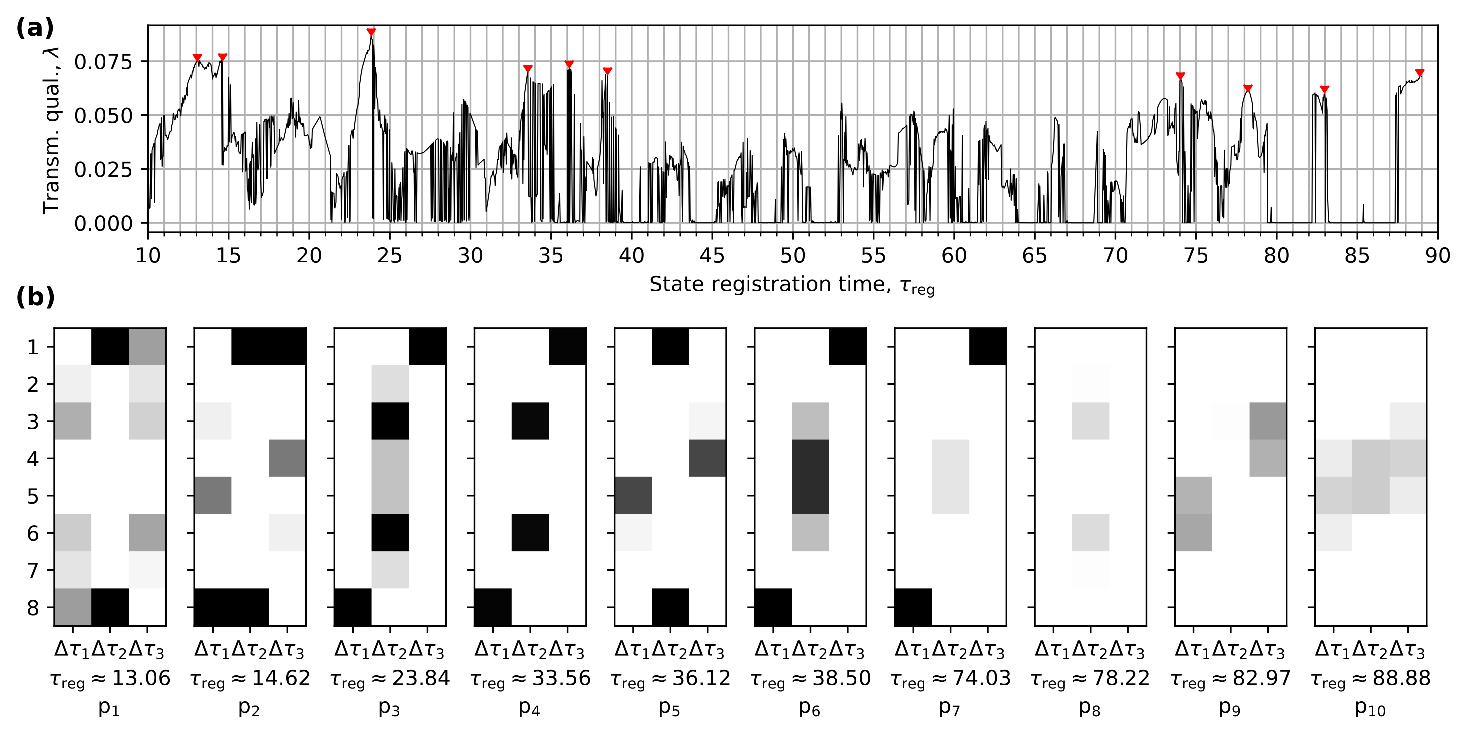}
  \caption{%
The state restoring in the chain of $N=8$ qubits. (a) The transmission quality~$\lambda$~(\ref{lammaxEx})
  as function of  the registration time-instant  $\tau_\mathrm{reg}$.
   The  ten best peaks are marked with the  red bullets.   (b) The {distributions of} damping rates  $\{a^{\mathrm{(opt)}}_{kj}\}$
    corresponding to the marked  peaks of the transmission quality~$\lambda$~(\ref{lammaxEx})
  }\label{fig:best-min-lambdas-over-transmission-time-n8}
\end{figure}
 Fig.~\ref{fig:best-min-lambdas-over-transmission-time-n8}a
shows the dependence of  $\lambda$~(\ref{lammax})
on the registration time-instant $\tau_\mathrm{reg}$.
Fig.~\ref{fig:best-min-lambdas-over-transmission-time-n8}b shows
the distributions of the damping rates $ \gamma^{\mathrm{(opt)}}=\{a^{\mathrm{(opt)}}_{kj}\}$ among different nodes (subscript $k$) and different time subintervals (subscript $j$) at specific time instants for state registration $\tau_{\mathrm{reg}}$ corresponding to the marked peaks on  Fig.~\ref{fig:best-min-lambdas-over-transmission-time-n8}a.
The white cells correspond to values $a^{\mathrm{(opt)}}_{kj}$  equal to zero,
while the completely black cells correspond to  $a^{\mathrm{(opt)}}_{kj}$  equal to one.
The transmission time instants $\tau_{\mathrm{reg}}$ correspond to the peak values in Fig.~\ref{fig:best-min-lambdas-over-transmission-time-n8}a.
The solution of restoring system~(\ref{U1002}) was  found with precision
 $\approx 10^{-8}$.

It can be seen from Fig.~\ref{fig:best-min-lambdas-over-transmission-time-n8}b
that almost all the presented {distributions}  of $\{ a^{\mathrm{(opt)}}_{kj}\}$  on plates ($p_1$) -- ($p_{10}$)  exhibit certain
 {\it Regularities:}
\begin{enumerate}
  \item \label{i1} All distributions demonstrate high central symmetry;
  \item\label{i2} Over the first time subinterval,
    there is no interaction between the environment and  sender,
    while over the last time subinterval there is no interaction between the environment and receiver (appropriate cells are white);
  \item The interaction of the environment {with a particular selected qubit   is switched on over at most a single subinterval (there is at most one colored cell in any horizontal triad of cells, accept, for instance, plates $p_1$ and $p_{10}$)};
  \item The two-qubit  sender and receiver  partially ($p_2$, $p_4$, $p_5$, $p_6$, $p_7$)  or fully  ($p_1$, $p_3$) participate in control, see
    Fig.~\ref{fig:best-min-lambdas-over-transmission-time-n8}b.
\end{enumerate}

\subsubsection{Centrally-symmetric distribution of damping rates}

In the above list of regularities, the first item is, perhaps, the most distinguished one.
It agrees with the concept that the symmetric transition line is most suitable for state transfer giving rise to the highest fidelity.
Therefore, we can expect that forcing  the symmetry in a damping rate distribution we achieve the best result.
For this reason, we imply this concept hereafter.
It is remarkable that the dependence of the transmission quality $\lambda$ on the registration time $\tau_{\mathrm reg}$ obtained using the symmetric setting  hardly differs from that shown in Fig.\ref{fig:best-min-lambdas-over-transmission-time-n8}a that justifies our assumption.
As an example, we apply the  centrally-symmetric concept to the optimization problem for the chain of 10 nodes ($N=10$).
The resulting transmission quality $\lambda$ as a function of $\tau_{\mathrm reg}$ is illustrated in
Fig.\ref{fig:best-min-lambdas-over-transmission-time-n10}a with the appropriate damping rate distributions in Fig.
\ref{fig:best-min-lambdas-over-transmission-time-n10}b, where the
 plates  $p_1$-$p_{10}$
are associated with the  transmission time-instances  $\tau_{\mathrm{reg}}$} corresponding to the marked peaks
in Fig.~\ref{fig:best-min-lambdas-over-transmission-time-n10}a.
Almost all distributions also inherit the above identified regularities (item n.1 is forced).

\begin{figure}[ht]
  \includegraphics[width=\textwidth]{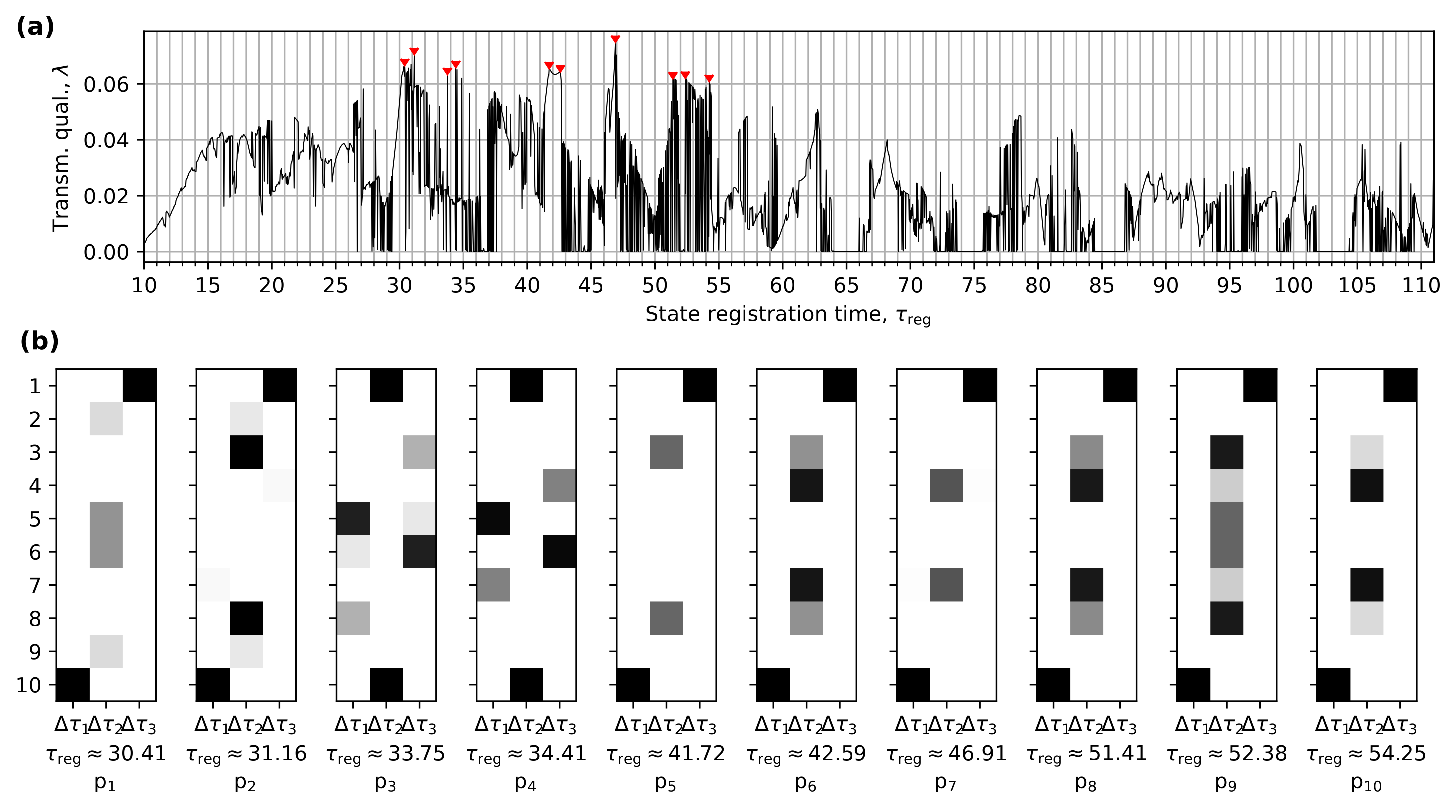}
  \caption{%
The state restoring in the chain of $N=10$ qubits.  (a) Transmission quality~$\lambda$~(\ref{lammaxEx})
    as a function of  registration time-instant $\tau_\mathrm{reg}$.
    The ten best peaks are marked with the red bullets. (b)  {The centrally-symmetric distributions of} damping rates $ \gamma^{\mathrm{(opt)}}$
    corresponding to the marked peaks of  transmission quality~$\lambda$~(\ref{lammaxEx}).
  }\label{fig:best-min-lambdas-over-transmission-time-n10}
\end{figure}
It can be seen in Figs.~\ref{fig:best-min-lambdas-over-transmission-time-n8}b
and~\ref{fig:best-min-lambdas-over-transmission-time-n10}b
that not all qubits are used for control.
In particular,
there is a sequence of well-isolated regions
that do not interact with the environment
and regions subjected to strong interaction.
Therefore, one can look for solutions with more specific patterns of
$\{a^{\mathrm{(opt)}}_{kj}\}$-distribution which are proposed below.

\subsubsection{Edges and center model}
\begin{figure}[ht]
  \includegraphics[width=\textwidth]{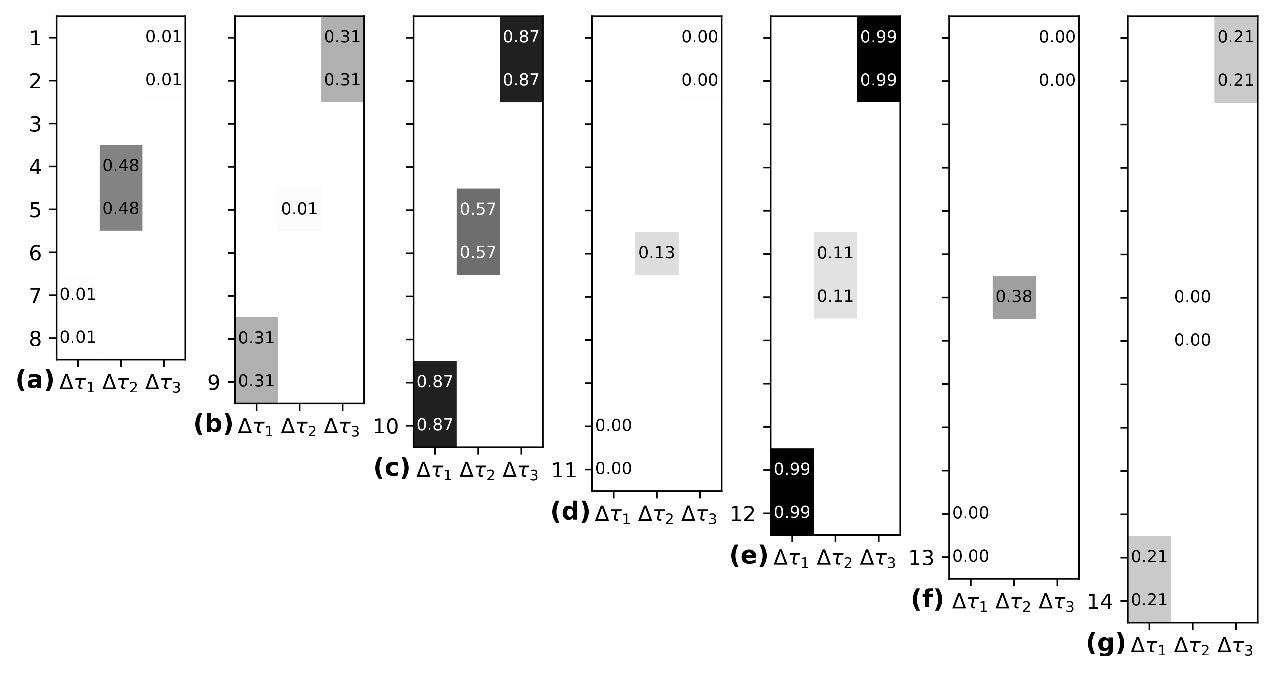}
  \caption{
    The damping rate values $ \gamma^{\mathrm{(opt)}}$
    corresponding to the maximum value of the transmission quality~$\lambda$~(\ref{lammaxEx})
    for different chain lengths $N$ and ``edges and center''-model,
    which is  the simplest  pattern of a damping rate distribution.
    The appropriate registration time instants are given in Table~\ref{tbl:lamndas-with-edges-and-center-pattern-by-length} together with the  transmission quality $\lambda$ and  maximal  absolute value of the $\lambda$-parameters $\lambda_\mathrm{max}$.
  }\label{fig:overview-best-min-lambdas-with-edges-and-center-pattern-by-length}
\end{figure}

The simplest template uses three small segments of the chain:
sender, receiver,
and one central qubit for odd-length chains or two central qubits for even-length chains.
We call this pattern as ``edges and center'' model.
The activation of control by each segment is switched on over different time subintervals.
The damping rates are all the same for each cell in the segment.
As a result, the control is performed using only two distinct values of damping rates.
The results of the best values of the smallest
 coefficient $\lambda$~(\ref{lammax})
obtained with this template for chains of different lengths are shown
in Fig.~\ref{fig:overview-best-min-lambdas-with-edges-and-center-pattern-by-length}.
The corresponding registration time instants $\tau_\mathrm{reg}$
and the
transmission qualities $\lambda$
%
are presented in Table~\ref{tbl:lamndas-with-edges-and-center-pattern-by-length} together with the appropriate  maximal absolute values of the $\lambda$-parameters $\lambda_\mathrm{max}$.

\begin{table}[ht]
  \begin{tabular}{|l|r|r|r|r|r|r|r|}
    \hline
    $N$ & 8 & 9 & 10 & 11 & 12 & 13 & 14 \\
    \hline
    $\tau_\mathrm{reg}$ & 75.218750 & 57.843750 & 30.687500 & 48.093750 & 34.656250 & 83.937500 & 73.625000 \\
    \hline
    $\lambda$ & 0.058715 & 0.033790 & 0.025932 & 0.038903 & 0.032684 & 0.033127 & 0.021490 \\
    \hline
    $\lambda_\mathrm{max}$ & 0.069925 & 0.084457 & 0.050267 & 0.111301 & 0.102252 & 0.085074 & 0.052726 \\
    \hline
  \end{tabular}
  \caption{%
    The registration time instant $\tau_\mathrm{reg}$, transmission quality $\lambda$ ~(\ref{lammaxEx})
   and appropriate maximal  absolute value of the $\lambda$-parameters $\lambda_\mathrm{max}$
    for different chain lengths $N$
    with the simplest pattern of controlled qubits called  ``edges and center'' model.
   The damping rate values $a^{\mathrm{(opt)}}_{kj}$  
    are shown in
    Fig.~\ref{fig:overview-best-min-lambdas-with-edges-and-center-pattern-by-length}.
  }\label{tbl:lamndas-with-edges-and-center-pattern-by-length}
\end{table}

\subsubsection{State restoring with equal $\lambda$-parameters}
In this section, we consider a particular case of restoring the 1-order coherence matrix when all $\lambda$ parameters equal each other.
Such restoring can be useful in certain applications
when we have to treat all restored matrix elements on equal footing.
Eqs.~(\ref{U1002}) remain the same,
while Eq.~(\ref{lammaxEx}) must be modified by the constraint $\lambda_{01}=\lambda_{10}$,
so that it gets the form
\begin{eqnarray}\label{single-lammaxEx}
  \lambda = \max_m \{ |\lambda_{01}(\tilde \gamma^{(m)} )| \}.
\end{eqnarray}
\begin{figure}[ht]
  \includegraphics[width=\textwidth]{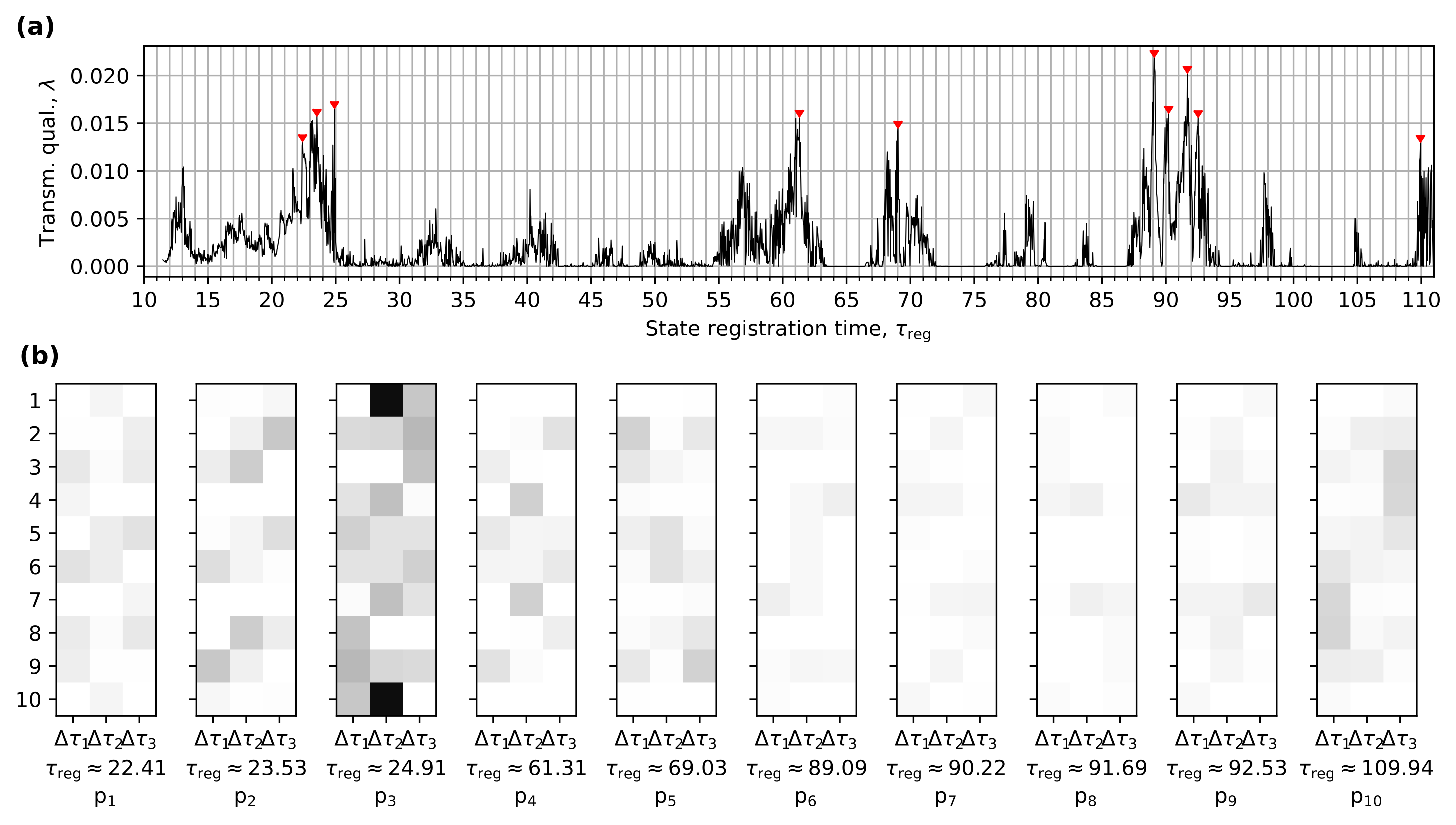}
  \caption{
State  restoring   in the chain of $N=10$ qubits with equal $\lambda$-parameters. (a)  The transmission quality~$\lambda$~(\ref{single-lammaxEx})
  as a function of   the registration time-instant  $\tau_\mathrm{reg}$.
    Ten best peaks are marked with the red bullets.
    (b)  The centrally-symmetric distributions of damping rates $\{a_{kj}^\mathrm{(opt)}\}$
    corresponding to the marked peaks of the transmission quality~$\lambda$~(\ref{single-lammaxEx})
  }\label{fig:best-min-single-lambdas-over-transmission-time-n10}
\end{figure}
Fig.~\ref{fig:best-min-single-lambdas-over-transmission-time-n10}a,
shows the dependence of $\lambda$~(\ref{single-lammaxEx})
on the registration time instant $\tau_\mathrm{reg}$.
Fig.~\ref{fig:best-min-single-lambdas-over-transmission-time-n10}b shows
the distributions of the damping rates $ \gamma^{\mathrm{(opt)}}=\{a^{\mathrm{(opt)}}_{kj}\}$ among different nodes (subscript $k$) and different time subintervals (subscript $j$) at specific time instants for state registration $\tau_{\mathrm{reg}}$
 corresponding to the marked peaks  in Fig.~\ref{fig:best-min-single-lambdas-over-transmission-time-n10}a.
It follows from Fig.~\ref{fig:best-min-single-lambdas-over-transmission-time-n10}b
that only  {\it Regularities}  n.\ref{i1} and  n.\ref{i2} of a damping rate distribution  defined above hold  in such case.

\section{Conclusions}
\label{Section:conclusions}
State-restoring protocols is a method of quantum information transfer that harmonize with the state-transfer and state-creation protocols. The quality of this method is characterized by the absolute values of scaling parameters
(or $\lambda$-parameters) that shrink the absolute values of the transferred elements of the density matrix. It is important that $\lambda$-parameters do not depend on the particular quantum state to be transferred and are defined only by the interaction Hamiltonian (or Lindbladian) and time instant for state registration, thus revealing their universality.

The state restoring protocol proposed in this paper uses a special time-dependent interaction with the 
environment thus replacing the restoring unitary transformation of the extended receiver that was originally 
used in state-restoring protocols \cite{FZ_2017,BFZ_Arch2018,Z_2018,FPZ_2021,BFLP_2022}. Although the 
$\lambda$-parameters are not large by absolute value for this type of restoring, the environment control might 
be combined with, for instance, the control by the local magnetic field \cite{FPZ_2024} and thus expand the 
freedom in choice of control parameters. { In addition, the absolute values of the $\lambda$-parameters 
can be increased using any special tool increasing the fidelity of state transfer, for instance, the week end-bonds 
\cite{GKMT,ABCVV,Z_2018}. However, even the absolute value $\sim 0.1$ obtainable in this paper can be 
acceptable, for instance, in the measurement based algorithm  for the perfect state 
transfer \cite{FWZ_arxive2025}.}

We also show the possibility to set all the $\lambda$-parameters equal to each other in restoring the 1-order coherence matrix, thus arranging the uniform compression of the transferred matrix elements. Although this step reduces the absolute value of $\lambda$-parameters in average, this kind of deformation of the sender initial state is valuable in the further development of the state-restoring protocols promoting the approach to the perfect state transfer.

{\bf Acknowledgments.}
The work was carried out as a part of a state task,
State Registration No. 124013000760-0. A.P. acknowledges partial support of the state task at MIAN.

\end{document}